\documentclass[10pt,a4paper]{article}
\usepackage[a4paper,margin=2cm]{geometry}

\usepackage{float}
\usepackage{graphicx}
\usepackage{amsmath}
\usepackage{mathtools}
\usepackage{amssymb}
\usepackage{grffile}
\usepackage{float}
\usepackage{xcolor}
\usepackage{gensymb}
\usepackage{hyperref}
\usepackage{dcolumn}
\usepackage{bm}
\usepackage{multirow}
\usepackage{makecell}
\usepackage{soul}
\usepackage{caption}
\usepackage{placeins}
\usepackage{bbm}
\usepackage{multicol}
\usepackage[super,sort&compress,comma]{natbib}

\vspace{-1.5em}
\title{\bf Conductance of silicon nanotube junctions \\in high magnetic fields}

\vspace{-1.0em}
\author{
Teresa Kulka$^1$,
Juan Alberto Canch\'e-Mart\'in$^{2,3}$,
Irina V. Lebedeva$^{2,4,5}$,\\
Jacek A. Majewski$^{1,6}$, 
Karolina Z. Milowska$^{*,7,8}$
}

\date{}

\begin{document}
\newgeometry{top=1.1cm,bottom=1.6cm,left=2cm,right=2cm}
\maketitle

\vspace{-2.0em}

\begin{center}
\footnotesize
$^1$ Faculty of Physics, University of Warsaw, Ludwika Pasteura 5, 02-093 Warsaw, Poland\\
$^2$ CIC nanoGUNE, Tolosa Hiribidea 76, Donostia-San Sebastian, 20018, Spain\\
$^3$ Autonomous University of Yucatan, Mexico\\
$^4$ Simune Atomistics, Donostia-San Sebastián, 20018, Spain\\
$^5$ Catalan Institute of Nanoscience and Nanotechnology (ICN2), CSIC and BIST, Campus UAB, Bellaterra, 08193 Barcelona, Spain\\
$^6$ Center for Terahertz Research and Applications - Centera 2, Centre for Advanced Materials and Technologies (CEZAMAT), Warsaw University of Technology, Warsaw, Poland\\
$^7$ BCMaterials, Basque Center for Materials, Applications and Nanostructures, UPV/EHU Science Park, 48940 Leioa, Spain\\
$^8$ Ikerbasque, Basque Foundation for Science, Plaza Euskadi, Bilbao, 48009, Spain\\
\end{center}
\begin{center}
\footnotesize
$^{*}$ E-mail: kz.milowska@bcmaterials.net
\end{center}

\vspace{-2.0em}

\begin{abstract}
We investigate coherent quantum transport through silicon nanotube (SiNT) junctions in high magnetic fields up to 60 T using a tight-binding model combined with the non-equilibrium Green's function formalism, and magnetic field included via Peierls substitution. We consider junctions of metallic nanotubes $(6,0)+(6,0)$ and semiconducting ones $(9,9)+(9,9)$, and examine the effects of the overlap length, inter-tube distance, magnetic-field direction, and field strength on the electronic transmission. In contrast to carbon nanotube junctions, the SiNT systems exhibit irregular transmission oscillations and do not show the emergence of highly conductive gateway states. The transmission is substantially more sensitive to a magnetic field perpendicular to the nanotube axis than to a parallel field, while increasing the field strength progressively modifies the transmission spectrum. Increasing the overlap length results in more frequent transmission oscillations, whereas increasing the inter-tube distance modifies their positions and amplitudes without changing their overall character. Generally, similar trends are observed for both types of junctions, involving metallic and semiconducting nanotubes. However, in the junction of semiconducting nanotubes, one observes peculiar additional field-dependent in-gap transmission features. These results demonstrate that the magnetic-field response of SiNT junctions is strongly governed by their geometry and differs qualitatively from that of pristine carbon nanotube junctions.
\end{abstract}

\begin{multicols}{2}

\section{Introduction}

The systems containing junctions and bundles of carbon nanotubes (CNTs) have been extensively researched in recent years. In particular, the issue of magneto-transport in such structures has been studied, mostly because of potential applications. In a series of papers, we investigated in detail, both experimentally and theoretically, the magnetotransport in the CNT systems \cite{Taming,Bulmer,manuscriptAGA}. These studies revealed that the magnetotransport in these systems is strongly influenced by quantum interference effects, even on the mesoscopic scale. The objective of the present paper is to investigate the extent to which the phenomena observed in carbon nanotube systems occur in silicon nanotube structures. Silicon nanotubes (SiNTs), the younger siblings of CNTs, are one-dimensional materials formed by rolling a silicene sheet into a cylindrical geometry \cite{Yang2005}, in a full analogy to CNTs. However, the main difference between the nanotube building blocks, namely graphene and silicene monolayers, is that graphene is rather flat, whereas the silicene layer exhibits considerable buckling,  involving the $sp^3$ character of bonding \cite{GuzmanVerri2007,Liu2021,Heidari2015} and the non-equivalence of the two Si atoms in the silicene unit cell. The electronic structure of SiNTs and their stability have been investigated in a series of papers on the {\it{ab-initio}} level of theory, mostly within the framework of density functional theory (DFT) \cite{Rashidi2024,Bai2004,Durgun2005,Wang2017}, as well as by employing the semi-empirical tight-binding (TB) computational scheme \cite{GuzmanVerri2007}. Also, the role of the external magnetic field in modifying various properties of SiNTs has been studied \cite{Konobeeva2018, Behzad2016, Chegel2024}.  In spite of the fairly voluminous literature on the properties of SiNTs, we are not aware of any studies involving the junctions of SiNTs in an external magnetic field and quantum magnetotransport in such structures.

Therefore, in this paper, we present coherent quantum transport calculations through model junctions composed of prototypical metallic (due to the $\pi^*$ and $\sigma^*$ mixing \cite{Yang2005}) (6,0) and semiconducting (9,9) SiNTs, placed under an external magnetic field oriented either parallel or perpendicular to the tube axis. Both tube geometries have been established as energetically stable \cite{Durgun2005}. The magnetic-field response in tubular geometries scales with the magnetic flux threaded through the cross-sectional area in the $yz$-plane \cite{Taming}, defined as the product of the tube diameter(7.44~\AA\, and 19.34 \AA, for (6,0) and (9,9), respectively) and the axial translation vector length  (6.63 \AA\, and 3.85 \AA, respectively). These specific chiralities were selected to ensure comparable unit-cell cross-sections. For the corresponding representative interfaces, the $(6,0)+(6,0)$ and $(9,9)+(9,9)$, we investigate how the junction geometry (specifically overlap length and inter-tube spacing) together with the magnetic field direction and strength determine the electronic transmission.

Unlike CNTs, SiNTs combine the geometric confinement characteristic of nanotubular systems with the rich electronic properties of silicon, while their curvature and structural reconstruction can strongly influence their electronic structure. Their diameter, chirality, and atomic configuration therefore provide important means of tuning their transport properties, making SiNTs promising candidates for investigating quantum confinement and nanoscale conduction.

\section{Computational procedure}

The large number of atoms in nanotube junctions makes direct  first-principles calculations of their electronic and transport  properties computationally demanding. We therefore describe the electronic structure of the SiNT junctions within an empirical  tight-binding (TB) framework, provided that the TB Hamiltonian  can be satisfactorily parametrised. This approach also enables an efficient non-perturbative treatment of external magnetic fields  through the Peierls substitution, as demonstrated in our previous studies of carbon nanotube junctions~\cite{Bulmer,Taming,manuscriptAGA}. 

The computational procedure consisted of two main stages. First,  the equilibrium geometries and electronic band structures of the  pristine $(6,0)$ and $(9,9)$ SiNTs were obtained within the DFT framework using the QuantumATK numerical package~\cite{Smidstrup2019,qATK}. In particular, structural  optimisation at the DFT level provides a reliable description of  the characteristic buckling of the SiNT walls. The resulting  electronic structures were subsequently used as reference data  for the parametrisation of the orthogonal four-orbital TB Hamiltonian employed in the present study. Details of the DFT  calculations are provided in the Supporting Information.

The subsequent construction of the SiNT junction devices and magnetotransport calculations was performed using our modified  implementations of the \texttt{sisl} Python library~\cite{sisl}, based on version 0.15.1, and \textsc{TBtrans}~\cite{tbtrans}, based on  \textsc{SIESTA} 5.2.0-alpha~\cite{Taming}. These modifications  enable the magnetic-field-induced complex Hamiltonian to be treated not only in the central scattering region but also in the  semi-infinite electrodes, allowing the Peierls phase to be included  consistently throughout the entire two-probe device. For each junction, \texttt{sisl} was used to construct the corresponding magnetic-field-dependent TB Hamiltonian, while the energy-resolved  transmission spectra were calculated within the non-equilibrium Green's function (NEGF) formalism using \textsc{TBtrans}. This procedure allows the transport properties to be calculated for magnetic fields up to 60\,T and for both parallel  and perpendicular field orientations.


\subsection{Construction of the tight-binding Hamiltonian}

 While the electronic structure of a CNT can be characterized with a good proximity by only considering $p_z$ orbitals \cite{Dresselhaus1996} in the TB Hamiltonian, in SiNTs, this approach works only for electron energies rather close to the Fermi level and for structures with sufficiently large diameter \cite{Liu2021}, \emph{i.e.}, starting from (18,0) for zigzag and from (15,15) for armchair SiNTs. The buckling present in SiNts enforces the need to use TB with four-orbitals, $s$, $p_x$, $p_y$, and $p_z$, per silicon atom \cite{Heidari2015}. The Hamiltonian can be written as
\begin{equation}
\mathbf{H}=\sum_{i}\epsilon_{i}a_i^\dagger a_i+
\sum_{\langle i,j\rangle}t_{ij}
\left(
a_i^\dagger a_j
+
a_j^\dagger a_i
\right)
\label{eq:TBhamiltonian}
\end{equation}
with $a_i^\dagger$ $(a_i)$ being the creation (annihilation) operators, $\epsilon_{i}$ on-site parameters, and $t_{ij}$ hopping parameters between orbitals $i$ and $j$, where $i = (n, \alpha)$, with $n$ being the lattice site index and $\alpha$ indicating one of the four orbital types. We have performed a survey of TB Hamiltonian parameters for silicon 1D and 2D structures through the existing literature and employed them to the calculation of the band structure of (6,0) and (9,9) SiNTs, however, we have not been able to reach a satisfactory agreement with DFT results. Therefore, starting from TB parameters of Grosso and Piermarocchi \cite{Grosso1995}, we established our own set of the orthogonal TB parameters: on-site $E_s$ = -4.0497 eV, $E_p$ = 1.0297 eV, for the nearest-neighbor (1) hopping $V_{ss\sigma1}$ = -0.2151 eV, $V_{sp\sigma1}$ = 7.6214 eV, $V_{pp\sigma1}$ = 6.6244 eV, $V_{pp\pi1}$ = -1.1324 eV; and for the next nearest neighbor (2) hopping $V_{ss\sigma2}$ = 0.8799 eV, $V_{sp\sigma2}$ = -0.1118 eV, $V_{pp\sigma2}$ = 2.1297 eV, $V_{pp\pi2}$ = -0.0001 eV. We also introduced the shift of on-site energies on the nonequivalent Si sites to be 0.4 eV \cite{Ahmadi2017}. The comparison between the band structures as obtained with TB and DFT methods are provided in Figs. S1 (a)-(b), and S2 (a)-(b) for (6,0) and (9,9) SiNTs, respectively. 

From the previously performed studies for carbon nanotube junctions \cite{Taming}, we know that the hopping between third nearest neighbors is also important, since it considerably influences the transmission function of the junctions. In the studied silicon systems, the distances $d$ between $3^{\text{rd}}$ NN lie in the range $4.10$ \AA\, $<d\leqslant$ $5.30$ \AA\. We assume that the hopping integrals between $3^{\text{rd}}$ NN are obtained from the ones for $2^{\text{rd}}$ NN employing the Harrison's scaling formula, $t^{(3)}_{nn'}= t^{(2)}_0 \left({d_0}/{d_{nn'}}\right)^2$, where $t^{(2)}_0$ is the reference hopping integral for the $2^{\text{rd}}$ NN, and the $d_0$ is the nominal second nearest neighbor distance  
(in the range of $2.35$ \AA\, $<d\leqslant$ $4.0$ \AA  in the investigated structures).


\subsection{Including magnetic field}
Because strong magnetic fields ($B$) must be treated non-perturbatively, the magnetic field was incorporated into the Hamiltonian via the Peierls substitution \cite{Peierls1933, SaitoDresselhaus1998}. It transforms the hopping parameters $t_{ij}$ of the Hamiltonian by adding a phase factor $\varphi$
\begin{equation}
    t_{ij}
    \;\rightarrow\; t_{ij}e^{i\varphi},
    \label{eq:peierlsH}
\end{equation}
\begin{equation}
    \varphi=\frac{2\pi}{\Phi_0}\int_{\mathbf r_i}^{\mathbf r_j} \mathbf A\cdot d\mathbf r,
    \label{eq:peierlsphase}
\end{equation}
with the coordinates at the atom $i$ being $\mathbf r_i=(x_i,y_i,z_i)$, $\mathbf A$ being the magnetic vector potential defined up to the gauge, and $\Phi_0=\frac{h}{e}$ the flux quantum with $h$ the Planck constant and $e$ the electron charge. The coordinate system is directed in a way so that electrons hop in the $y$ direction between separate SiNTs forming a junction, see Fig.~\ref{fig:60}(a,b) and Fig.~\ref{fig:99}(a,b). We set $B_y=0$, because the magnetic field parallel to the hopping direction ($B_y\neq0$) would not induce the Aharonov-Bohm phases \cite{PhysRev.115.485} between alternative electron pathways and, what follows, would not change transmission. Therefore, we research only systems in $B_x$ (perpendicular) and $B_z$ (parallel) magnetic fields. The gauge is chosen so it preserves translational invariance along $z$ direction, with the vector potential $\mathbf A = (-yB_z,\, 0,\, y B_x)$, and the Peierls phase being
\begin{equation}
    \varphi=\frac{\pi}{\Phi_0}[B_x(z_j - z_i)-B_z(x_j - x_i)](y_i + y_j).
    \label{eq:phase-f}
\end{equation}
In the present calculations, the Zeeman contribution is neglected,  as we focus on the orbital response of the SiNT junctions to the  external magnetic field. For electrons with a $g$-factor close to 2,  the Zeeman splitting at the maximum field considered here, B=60\,T, is approximately 7\,meV. In our previous study of CNT  junctions~\cite{Taming}, the influence of the Zeeman term was examined  explicitly and was found to mainly introduce two independently energy-shifted spin contributions to the transmission, without changing the underlying orbital-interference mechanism.  In this way, the magnetic-field-dependent TB Hamiltonian is created, providing a solid foundation for the subsequent calculation of the electronic transport properties.

\subsection{Transport calculations}
The transport calculations were performed using non-equilibrium Green’s function (NEGF) formalism \cite{Papior2017, PhysRevB.65.165401}. The junction device can be divided into three regions: left electrode $L$, scattering (central) region $C$, and right electrode $R$. Transmission amplitude of an electron with energy $E$ between two electrodes $L$ and $R$ is computed from the Landauer–Büttiker formula
\begin{equation}
    T_{L\rightarrow R}(E)
    = \mathrm{Tr}\!\left[\mathbf \Gamma_R(E)\, \mathbf G_C(E)\, \mathbf \Gamma_L(E)\, \mathbf G_C^\dagger(E)\right],
    \label{eq:transmission}
\end{equation}
where $\mathbf \Gamma_{\alpha},\, \alpha=L,R$ is a matrix describing the coupling to the lead $L/R$
\begin{equation}
    \mathbf \Gamma_{\alpha}(E)=i[\mathbf \Sigma_{\alpha}(E)-\mathbf \Sigma_{\alpha}^\dagger(E)],
    \label{eq:scattering}
\end{equation}
and $\mathbf G_C$ is a central region (dressed by self-energies) Green's function expressed in terms of its Hamiltonian $\mathbf H_C$, its overlap matrix $\mathbf S_C$ (equal to $\mathbf{1}$ for our orthogonal basis set), and electrodes' self-energies $\mathbf \Sigma_{\alpha},\, \alpha=L,R$, in the limit $\eta\rightarrow0^+$
\begin{equation}
    \mathbf G_C^{-1}(E)= (E+i\eta) \mathbf S_C - \mathbf H_C - \mathbf \Sigma_L(E)-\mathbf \Sigma_R(E).
    \label{eq:G}
\end{equation}
$\mathbf G_C$ is dressed by self-energies, \emph{i.e.} self-energies account for the effect of the semi-infinite electrodes on the finite scattering region and arise from integrating out the leads' degrees of freedom. Electrode's $\alpha=L,R$ self-energy is
\begin{equation}
    \mathbf \Sigma_{\alpha}(E)=(\mathbf H_{C\alpha}-E \mathbf S_{C\alpha}) \mathbf G_{\alpha}(E)(\mathbf H_{C\alpha}^{\dagger}-E \mathbf S_{C\alpha}^{\dagger})
    \label{eq:sigma}
\end{equation}
with its Green's function
\begin{equation}
    \mathbf G_{\alpha}^{-1}(E)= (E+i\eta) \mathbf S_{\alpha} - \mathbf H_{\alpha}.
    \label{eq:Glead}
\end{equation}
The device Hamiltonian $\mathbf H$ and overlap $\mathbf S$ matrices are
\begin{equation}
\mathbf H=
\begin{pmatrix}
\mathbf H_C & \mathbf H_{C\alpha} \\
\mathbf H_{\alpha C} & \mathbf H_{\alpha}
\end{pmatrix},\,
\mathbf S=
\begin{pmatrix}
\mathbf S_C & \mathbf S_{C\alpha} \\
\mathbf S_{\alpha C} & \mathbf S_{\alpha}
\end{pmatrix},\,
\alpha=L,R,
\end{equation}
where $\mathbf H_C$/$\mathbf H_{\alpha}$ and $\mathbf S_C$/$\mathbf S_{\alpha}$ are Hamiltonian and overlap matrices of the scattering region/semi-infinite electrodes, with $\mathbf H_{\alpha C}=\mathbf H_{C\alpha}^{\dagger}$ and $\mathbf S_{\alpha C}=\mathbf S_{C\alpha}^{\dagger}$ being Hermitian matrices describing the coupling between central region and electrodes. The magnetic field changes the Hamiltonian by $\delta \mathbf H$ and self-energies by $\delta \mathbf \Sigma$ by the Peierls substitution, which can be incorporated into the Green's function as
\begin{equation}
    \mathbf G_C^{-1}(E)= (E+i\eta) \mathbf S_C - \mathbf H_C - \delta \mathbf H - \mathbf \Sigma_L(E) - \mathbf \Sigma_R(E) - \delta \mathbf \Sigma.
    \label{eq:dG}
\end{equation}

\section{Results and discussion}

\begin{figure*}
     \centering
     \includegraphics[width=1.0\linewidth]{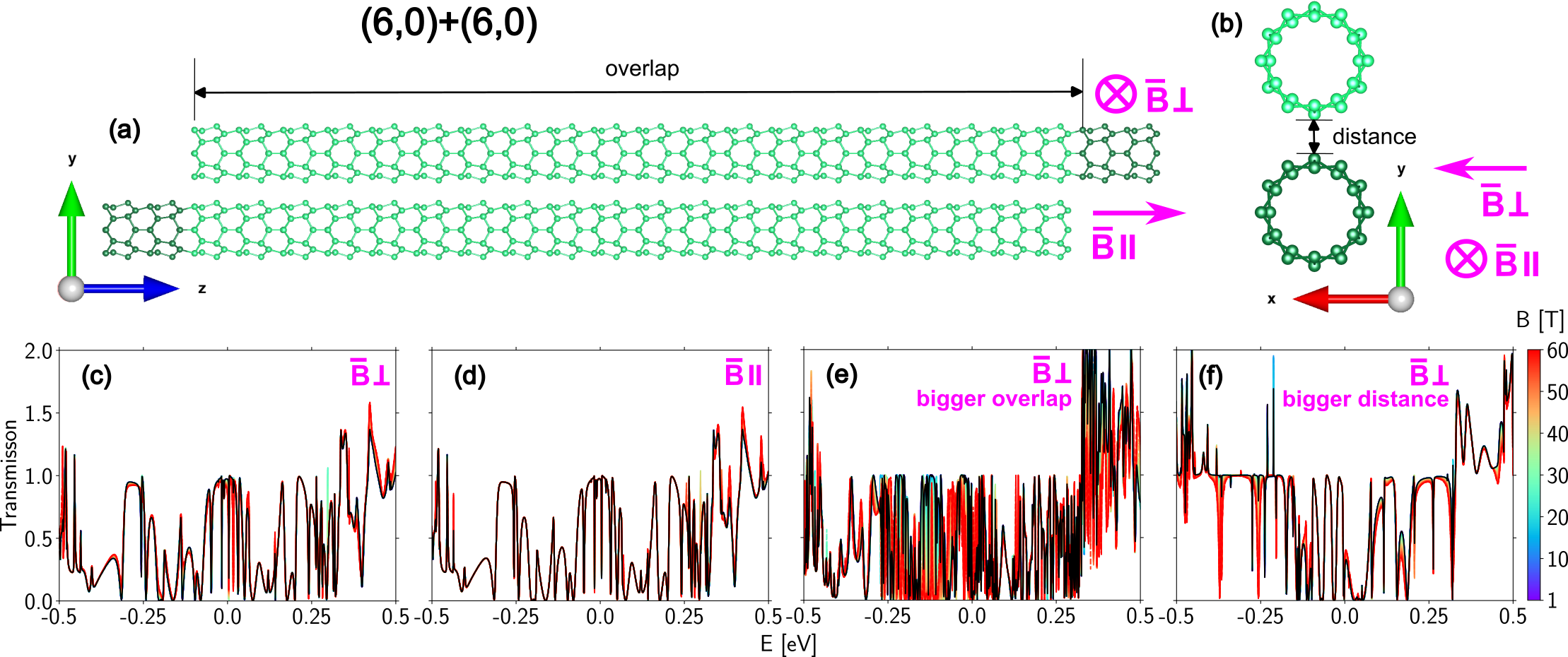}
     \caption{(a) Side and (b) cross-sectional views of the metallic $(6,0)+(6,0)$ SiNT junction. Each tube (light green) is connected to a semi-infinite electrode of the same chirality (dark green). The overlap length and inter-tube distance are indicated, as well as the directions corresponding to perpendicular and parallel magnetic fields. (c-f) Transmission spectra of the $(6,0)+(6,0)$ SiNT junction at zero magnetic field (black lines) with an overlap lengths and inter-tube distances (c,d) 13.26 nm (20 units) and 0.40 nm, (e) 53.04 nm (80 units) and 0.40 nm, (f) 13.26 nm (20 units) and 0.45 nm, respectively. Coloured lines correspond to (c,e,f) perpendicular and (d) parallel magnetic fields up to 60 T.}
     \label{fig:60}
\end{figure*}

\begin{figure*}
     \centering
     \includegraphics[width=0.99\linewidth]{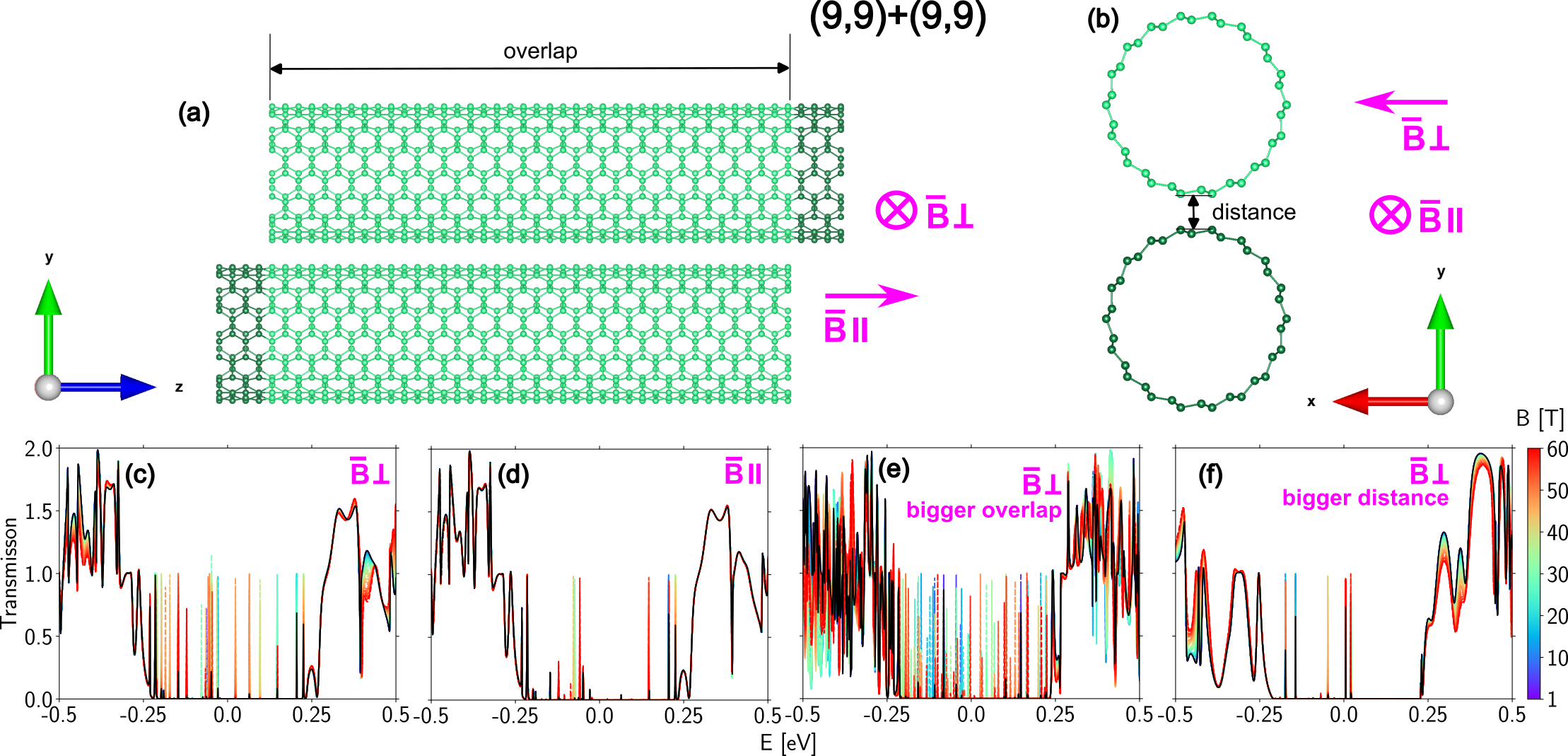}
     \caption{(a) Side and (b) cross-sectional views of the semiconducting $(9,9)+(9,9)$ SiNT junction. Each tube (light green) is connected to a semi-infinite electrode of the same chirality (dark green). The overlap length and inter-tube distance are indicated, as well as the directions corresponding to perpendicular and parallel magnetic fields. (c-f) Transmission spectra of the $(9,9)+(9,9)$ SiNT junction at zero magnetic field (black lines) with an overlap lengths and inter-tube distances (c,d) 7.69 nm (20 units) and 0.40 nm, (e) 30.76 nm (80 units) and 0.40 nm, (f) 7.69 nm (20 units) and 0.45 nm, respectively. Coloured lines correspond to (c,e,f) perpendicular and (d) parallel magnetic fields up to 60 T.}
     \label{fig:99}
\end{figure*}


The considered junction geometries are shown in Figs.~\ref{fig:60}(a,b) and~\ref{fig:99}(a,b). In each case, the two nanotubes have the same chirality and are connected to semi-infinite electrodes of the corresponding chirality. The transmission spectra obtained for different magnetic field directions, overlap lengths, and inter-tube distances are presented in Figs.~\ref{fig:60}(c-f) and~\ref{fig:99}(c-f), while for a bigger energy range in Figs.~\ref{fig:fulltrans60}(a-d) and~\ref{fig:fulltrans99}(a-d).

In contrast to the pronounced and relatively regular transmission oscillations observed for the CNT junctions~\cite{Taming}, at zero magnetic field (black lines in Figs.~\ref{fig:60}(c-f) and~\ref{fig:99}(c-f)), the SiNT junctions exhibit a more irregular oscillatory structure. This behavior is visible for both the metallic $(6,0)+(6,0)$ and semiconducting $(9,9)+(9,9)$ junctions. Moreover, the magnetic field does not give rise to the highly conductive gateway states observed previously in CNT junctions~\cite{Taming}. Instead, the field mainly modifies the existing irregular transmission features. The overall magnetic-field response of the SiNT junctions is also weaker than that found for the corresponding CNT systems~\cite{Taming}. The possible explanation is the buckled structure of the SiNTs, which modifies the orbital geometry~\cite{Liu2021}, and consequently the magnetic-field-dependent Peierls phases associated with the hopping pathways~\cite{Behzad2016}.

When considering the effect of the magnetic-field direction, for both SiNT junctions, the transmission is considerably less sensitive to a magnetic field parallel to the nanotube axis than to a perpendicular field, compare Figs.~\ref{fig:60}(c) and~\ref{fig:60}(d), and similarly Figs.~\ref{fig:99}(c) and~\ref{fig:99}(d). This indicates that, for the junction geometries investigated here, the orbital response is strongly dependent on the magnetic field orientation. While the perpendicular field produces visible changes in the transmission spectrum, the modifications induced by the parallel field remain comparatively weak over the considered experimentally available field range, \emph{i.e.}, up to 60 T. Therefore, we further focus primarily on the perpendicular magnetic field.

The overlap length has a pronounced effect on the transmission. Increasing the overlap 4 times, \emph{i.e.}, from 13.26~nm to 53.04~nm for the $(6,0)+(6,0)$ junction, see Figs.~\ref{fig:60}(c) and~\ref{fig:60}(e), and from 7.69~nm to 30.76~nm for the $(9,9)+(9,9)$ junction, see Figs.~\ref{fig:99}(c) and~\ref{fig:99}(e), leads to more frequent irregular oscillations in the transmission spectrum. Although these oscillations in SiNT junctions do not exhibit the regularity found in CNTs~\cite{Taming}, their frequency is still strongly controlled by the junction overlap length. This behavior is consistent with the increased number of interfering propagation pathways available for longer overlap regions.

The inter-tube distance also modifies the band structure, and hence transmission spectrum. The band structures of 2 units cells for inter-tube distances of 0.40~nm and 0.45~nm are depicted in Fig.~\ref{fig:bands60}(c,d) for $(6,0)+(6,0)$ and Fig.~\ref{fig:bands99}(c,d) for $(9,9)+(9,9)$ SiNT junctions. The inter-tube distances were taken a bit bigger than Si--Si distance between layers of multilayer silicene \cite{DePadova2017}. Increasing the separation from 0.40~nm to 0.45~nm, see Figs.~\ref{fig:60}(c) and Figs.~\ref{fig:99}(c) versus Figs.~\ref{fig:60}(f) and Figs.~\ref{fig:99}(f), also changes the positions and amplitudes of individual transmission features. Nevertheless, the overall structure of the spectrum remains qualitatively similar. This indicates that the inter-tube distance affects the magnitude of the inter-tube coupling, while the characteristic irregular structure of the transmission is preserved. As a result, the effect is distinct from a qualitative change of the transport mechanism.

As expected, increasing the magnetic-field strength leads to progressively larger modifications of the transmission. The deviations from the zero-field spectrum increase with increasing field strength, with the most pronounced changes occurring at the highest fields considered, up to 60~T. This behavior is observed for both SiNT chiralities and for the different junction geometries shown in Figs.~\ref{fig:60}(c-f) and~\ref{fig:99}(c-f). Therefore, the response evolves continuously with the applied field rather than appearing only above a characteristic field threshold.

Interestingly, the transmission spectra obtained for the SiNT junctions bear a qualitative resemblance to those of defected CNT junctions reported previously~\cite{manuscriptAGA}. In both cases, the transmission is characterized by irregular rather than highly regular oscillatory features, and the magnetic field primarily modifies these features without producing the pronounced high-conductance gateway states found in pristine CNT junctions~\cite{Taming}. This similarity suggests that buckling can play a role comparable to that of structural disorder or defects in determining the interference pattern.

The comparison between the ($(6,0)+(6,0)$ and $(9,9)+(9,9)$ SiNT junctions further shows that the main trends are not restricted to the metallic system. The semiconducting $(9,9)+(9,9)$ junction exhibits the same qualitative dependence on magnetic-field direction, field strength, overlap length, and inter-tube distance as the $(6,0)+(6,0)$ junction. This suggests that the observed magnetic-field response is primarily governed by the junction geometry and inter-tube coupling rather than by the metallic or semiconducting character. At the same time, the $(9,9)+(9,9)$ junction displays additional transmission peaks within the band gap region, see Fig.~\ref{fig:99}(c-f). These field-dependent in-gap peaks are more frequent for bigger overlap length and in perpendicular field than parallel, providing an additional way through which the magnetic field can modify transport in the semiconducting system.

\section{Conclusions}

We have investigated coherent quantum transport through metallic $(6,0)+(6,0)$ and semiconducting $(9,9)+(9,9)$ SiNT junctions in magnetic fields up to 60~T. In both systems, the transmission is considerably more sensitive to the magnetic field perpendicular to the nanotube axis than to the parallel field. Increasing the overlap length results in more frequent irregular transmission oscillations, while increasing the inter-tube distance modifies their positions and amplitudes without changing the overall character of the transmission spectrum.

The SiNT junctions differ from the pristine CNT junctions studied previously, as they exhibit irregular rather than pronounced regular transmission oscillations and do not show the emergence of highly conductive gateway states. Their response to the magnetic field is also weaker, which may be related to the buckled atomic structure of SiNTs. Despite these differences, the same main trends are observed for the $(6,0)+(6,0)$ and $(9,9)+(9,9)$ junctions composed of the metallic and the semiconducting nanotubes, respectively.  This indicates that the response of the junctions to the magnetic field cannot be ascribed to the metallic character of nanotubes constituting the junction.  The junction of the semiconducting nanotubes additionally exhibits field-dependent transmission features within the band-gap region. Overall, the presented results establish the main characteristics of magnetotransport through SiNT junctions, provide a basis for comparison with the corresponding carbon structures, and might invoke interest in experimental studies of silicon nanotube systems.

\section*{Data availability}
The data supporting the findings of this study, including the processed transmission spectra and structural models used in the calculations, are  available from the corresponding authors upon reasonable request. The  modified versions of the \texttt{sisl} and \textsc{TBtrans} codes used for  the magnetic-field transport calculations are also available from the corresponding authors upon reasonable request, subject to the licensing  terms of the original \texttt{sisl} and \textsc{TBtrans} packages. Additional  methodological details, band structures, and transmission spectra are provided in the Supporting Information (SI), available at DOI:...

\section*{Author contributions}
The theoretical concept of the work was developed jointly by J. A. M., K. Z. M. and T. K. 
T. K. performed the numerical simulations under the supervision of J. A. M. and K. Z. M. and contributed to the development of the sisl implementation required for magnetic-field simulations. J. A. C. M. established the tight-binding parameters used for the calculations. T. K., J. A. M. and K. Z. M. contributed to the interpretation of the theoretical results. I. V. L. contributed to the development of the SIESTA implementation required for magnetic-field simulations. K. Z. M. performed the DFT structural optimisations in QuantumATK and secured funding. T. K. prepared the first version of the manuscript, which was subsequently revised by J. A. M., J. A. C. M., I. V. L. and K. Z. M. All authors discussed the results and contributed to the final manuscript.

\section*{Acknowledgements}
T. K. and K. Z. M. gratefully acknowledge the Interdisciplinary Centre for Mathematical and Computational Modelling at the University of Warsaw, Poland (Grant No. G47-5) for providing computer facilities and technical support. T. K., K. Z. M. and I. V. L. also acknowledge the technical and human support provided by the DIPC Supercomputing Center, Spain. T. K. and K. Z. M. gratefully acknowledge the Agencia Estatal de Investigación, Ministerio de Ciencia e Innovación, Spain, for funding this research through the Proyectos de Generación de Conocimiento 2022 programme (PID2022-139776NB-C65) and the Proyectos de Generación de Conocimiento 2025 programme (PID2025-174822NB-C32). K. Z. M. also would like to thank the European Commission (Marie SklodowskaCurie Cofund Programme; Grant No. H2020-MSCA-COFUND-2020-101034228-WOLFRAM2) for supporting this research.  K.Z.M also would like to aknowledge the support from the Ramón y Cajal grant RYC2024-051436-I, funded by MICIU/AEI/10.13039/501100011033 and by ESF+. J. A. M. acknowledges the support from Centera2 project (FENG.02.02-IP.02.01-IP.05-T0004/23) funded with IRA FENG program of Foundation for Polish Science, and cofinanced by the EU FENG Programme. I. V. L. acknowledges support from the EuroHPC JU under the MAX (Materials design at the Exascale) project (Grant No. 101093374), and from the Spanish MCIN/AEI/10.13039/501100011033 and the European Union NextGenerationEU/PRTR through Grant No. PCI2022-134972-2. ICN2 is supported by the CERCA programme (Generalitat de Catalunya) and the Severo Ochoa Centres of Excellence programme (Grant No. CEX2021-001214-S), funded by MCIN/AEI/10.13039/501100011033.

\end{multicols}

\bibliographystyle{rsc}
\bibliography{biblio}

@book{Dresselhaus1996,
  title     = {Science of Fullerenes and Carbon Nanotubes},
  author    = {Dresselhaus, M. S. and Dresselhaus, G. and Eklund, P. C.},
  publisher = {Academic Press},
  year      = {1996},
  address   = {San Diego},
  isbn      = {9780122218200}
}

@article{Behzad2016,
  author  = {Behzad, Somayeh and Chegel, Raad},
  title   = {Magnetic Field-Induced Splitting of Optical Spectra in Silicon Nanotubes: Tight Binding Calculations},
  journal = {Silicon},
  volume  = {8},
  number  = {1},
  pages   = {43--55},
  year    = {2016},
  doi     = {10.1007/s12633-014-9230-2}
}

@article{Konobeeva2018,
  author  = {Konobeeva, N. N. and Evdokimov, R. A. and Belonenko, M. B.},
  title   = {Magnetic Field Effect on Ultrashort Two-dimensional Optical Pulse Propagation in Silicon Nanotubes},
  journal = {Russian Physics Journal},
  volume  = {61},
  number  = {1},
  pages   = {157--161},
  year    = {2018},
  doi     = {10.1007/s11182-018-1379-5}
}

@article{Chegel2024,
  author  = {Chegel, Raad},
  title   = {Comparative study of third harmonic generation in carbon and silicene nanotubes under magnetic fields},
  journal = {Scientific Reports},
  volume  = {14},
  pages   = {31227},
  year    = {2024},
  doi     = {10.1038/s41598-024-82561-x}
}

@article{Liu2021,
  author  = {Liu, Hsin-Yi and Lin, Ming-Fa and Wu, Jhao-Ying},
  title   = {Essential Electronic Properties of Silicon Nanotubes},
  journal = {Nanomaterials},
  volume  = {11},
  number  = {10},
  pages   = {2475},
  year    = {2021},
  doi     = {10.3390/nano11102475}
}

@article{Grosso1995,
  author  = {Grosso, G. and Piermarocchi, C.},
  title   = {Tight-binding model and interactions scaling laws for silicon and germanium},
  journal = {Physical Review B},
  volume  = {51},
  number  = {23},
  pages   = {16772--16777},
  year    = {1995},
  doi     = {10.1103/PhysRevB.51.16772}
}

@article{Peierls1933,
  author  = {Rudolf E. Peierls},
  title   = {{Zur Theorie des Diamagnetismus von Leitungselektronen}},
  journal = {Zeitschrift f\"ur Physik},
  volume  = {80},
  pages   = {763--791},
  year    = {1933},
  doi     = {10.1007/BF01342591},
}

@book{SaitoDresselhaus1998,
  author    = {R. Saito and G. Dresselhaus and M. S. Dresselhaus},
  title     = {{Physical Properties of Carbon Nanotubes}},
  publisher = {Imperial College Press},
  year      = {1998},
  address   = {London},
  isbn      = {9781860942235},
}

@article{PhysRev.115.485,
  title = {Significance of Electromagnetic Potentials in the Quantum Theory},
  author = {Aharonov, Y. and Bohm, D.},
  journal = {Phys. Rev.},
  volume = {115},
  pages = {485--491},
  numpages = {0},
  year = {1959},
  doi = {10.1103/PhysRev.115.485},
}

@article{PhysRevB.65.165401,
  title = {Density-functional method for nonequilibrium electron transport},
  author = {Brandbyge, Mads and Mozos, Jos\'e-Luis and Ordej\'on, Pablo and Taylor, Jeremy and Stokbro, Kurt},
  journal = {Phys. Rev. B},
  volume = {65},
  issue = {16},
  pages = {165401},
  numpages = {17},
  year = {2002},
  month = {Mar},
  publisher = {American Physical Society},
  doi = {10.1103/PhysRevB.65.165401},
  url = {https://link.aps.org/doi/10.1103/PhysRevB.65.165401}
}

@ARTICLE{Papior2017,
  author = {Papior, Nick and Lorente, Nicol{\'a}s and Frederiksen, Thomas and
    Garc{\'i}a, Alberto and Brandbyge, Mads},
  title = {{I}mprovements on non-equilibrium and transport {G}reen function
    techniques: {T}he next-generation {T}ran{SIESTA}},
  journal = {Comput. Phys. Commun.},
  year = {2017},
  volume = {212},
  pages = {8--24},
  doi = {10.1016/j.cpc.2016.09.022},
}

@article{Taming,
  author    = {Kulka, T. and Lekawa-Raus, A. E. and Bulmer, J. S. and Koziol, K. and Balakirev, F. F. and Lebedeva, I. V. and Majewski, J. A. and Marganska, M. and Milowska, K. Z.},
  title     = {Taming quantum interference: a route to high electrical conductance in carbon nanotube assemblies},
  journal   = {Nanoscale},
  year      = {2026},
  doi       = {10.1039/D6NR02332K}
}

@unpublished{manuscriptAGA,
  author = {Lekawa-Raus, A. E. and Bulmer, J. S. and Kulka, T. and Marganska, M. and Papior, N. and  Rickel, D. G and Balakirev, F. F. and Majewski, J. A. and Koziol, K. and Milowska, K. Z.},
  title  = {Quantum Limits of Electronic Transport in
Nanostructured Macroscopic Conductors},
  note   = {arXiv preprint},
  year   = {2026},
  doi    = {arXiv.2605.02295},
}

@article{GuzmanVerri2007,
  author  = {Guzm{\'a}n-Verri, G. G. and Lew Yan Voon, L. C.},
  title   = {Electronic structure of silicon-based nanostructures},
  journal = {Physical Review B},
  volume  = {76},
  number  = {7},
  pages   = {075131},
  year    = {2007},
  doi     = {10.1103/PhysRevB.76.075131}
}

@article{Bulmer,
  title = {Competing conduction mechanisms in high performance carbon nanotube fibers},
  author = {John Bulmer and Chris Kovacs and Thomas Bullard and Charlie Ebbing and Timothy Haugan and Ganesh Pokharel and Stephen D. Wilson and Fedor F. Balakirev and Oscar A. Valenzuela and Michael A. Susner and David Turner and Pengyu Fu and Teresa Kulka and Jacek Majewski and Irina Lebedeva and Karolina Z. Milowska and Agnieszka Lekawa-Raus and Magdalena Marganska},
  journal = {Carbon},
  volume = {248},
  pages = {121162},
  year = {2026},
  month = {Feb},
  publisher = {Elsevier},
  doi = {10.1016/j.carbon.2025.121162},
  url = {https://doi.org/10.1016/j.carbon.2025.121162}
}

@article{Heidari2015,
  author  = {Heidari, Mohsen and Ahmadi, Vahid and Darbari, Sara},
  title   = {Electronic and Optical Properties of Single-Walled Silicon Nanotubes: The {sp3} Tight-Binding Model},
  journal = {Modares Journal of Electrical Engineering},
  volume  = {15},
  number  = {3},
  pages   = {27--31},
  year    = {2015},
  url     = {https://mjee.modares.ac.ir/}
}

@article{Yang2005,
  author  = {Yang, Xiaobao and Ni, Jun},
  title   = {Electronic Properties of Single-Walled Silicon Nanotubes Compared to Carbon Nanotubes},
  journal = {Physical Review B},
  volume  = {72},
  number  = {19},
  pages   = {195426},
  year    = {2005},
  doi     = {10.1103/PhysRevB.72.195426},
  url     = {https://doi.org/10.1103/PhysRevB.72.195426}
}

@article{Rashidi2024,
  author  = {Rashidi, Donna and Hakimi, Maryam and Frank, Irmgard and Nadimi, Ebrahim},
  title   = {Exploring the Structural and Electronic Properties of Different Types of Silicon Nanotubes: A First-Principles Study},
  journal = {ACS Applied Electronic Materials},
  volume  = {6},
  number  = {10},
  pages   = {7540--7550},
  year    = {2024},
  doi     = {10.1021/acsaelm.4c01372},
  url     = {https://doi.org/10.1021/acsaelm.4c01372}
}

@article{Bai2004,
  author  = {Bai, Jaeil and Zeng, X. C. and Tanaka, Hideki and Zeng, J. Y.},
  title   = {Metallic Single-Walled Silicon Nanotubes},
  journal = {Proceedings of the National Academy of Sciences},
  volume  = {101},
  number  = {9},
  pages   = {2664--2668},
  year    = {2004},
  doi     = {10.1073/pnas.0308467101},
  url     = {https://doi.org/10.1073/pnas.0308467101}
}

@article{Ahmadi2017,
  author  = {Ahmadi, Neda and Shokri, A. A.},
  title   = {Optoelectronic Properties of Silicon Hexagonal Nanotubes under an Axial Magnetic Field},
  journal = {Optics Communications},
  volume  = {395},
  pages   = {282--288},
  year    = {2017},
  doi     = {10.1016/j.optcom.2016.07.049},
  url     = {https://doi.org/10.1016/j.optcom.2016.07.049}
}

@article{DePadova2017,
  author  = {De Padova, Paola and Feng, Haifeng and Zhuang, Jincheng and Li, Zhi and Generosi, Amanda and Paci, Barbara and Ottaviani, Carlo and Quaresima, Claudio and Olivieri, Bruno and Krawiec, Mariusz and Du, Yi},
  title   = {Synthesis of Multilayer Silicene on {Si}(111)$\sqrt{3} \times \sqrt{3}$-{Ag}},
  journal = {The Journal of Physical Chemistry C},
  volume  = {121},
  number  = {48},
  pages   = {27182--27190},
  year    = {2017},
  doi     = {10.1021/acs.jpcc.7b09286},
  url     = {https://doi.org/10.1021/acs.jpcc.7b09286}
}

@article{Durgun2005,
  author  = {Durgun, E. and Tongay, S. and Ciraci, S.},
  title   = {Silicon and {III-V} Compound Nanotubes: Structural and Electronic Properties},
  journal = {Physical Review B},
  volume  = {72},
  number  = {7},
  pages   = {075420},
  year    = {2005},
  doi     = {10.1103/PhysRevB.72.075420},
  url     = {https://doi.org/10.1103/PhysRevB.72.075420}
}

@article{Wang2017,
  author  = {Wang, Chongze and Fu, Xiaonan and Guo, Yangyang and Guo, Zhengxiao and Xia, Congxin and Jia, Yu},
  title   = {Band Gap Scaling Laws in Group {IV} Nanotubes},
  journal = {Nanotechnology},
  volume  = {28},
  number  = {11},
  pages   = {115202},
  year    = {2017},
  doi     = {10.1088/1361-6528/aa5b3e},
  url     = {https://doi.org/10.1088/1361-6528/aa5b3e}
}

@article{Wang2021,
  author  = {Wang, Zifeng and Ye, Shizhuo and Wang, Hao and He, Jin and Huang, Qijun and Chang, Sheng},
  title   = {Machine learning method for tight-binding Hamiltonian parameterization from ab-initio band structure},
  journal = {npj Computational Materials},
  volume  = {7},
  pages   = {11},
  year    = {2021},
  doi     = {10.1038/s41524-020-00490-5}
}

@article{Smidstrup2019,
  title = {QuantumATK: An integrated platform of electronic and atomic-scale modelling tools},
  author = {Smidstrup, S. and Markussen, T. and Vancraeyveld, P. and Wellendorff, J. and Schneider, J. and Gunst, T. and Verstichel, B. and Stradi, D. and Khomyakov, P. A. and Vej-Hansen, U. G. and others},
  journal = {Journal of Physics: Condensed Matter},
  volume = {32},
  pages = {015901},
  year = {2019},
  doi = {10.1088/1361-648X/ab4007}
}

@misc{qATK,
  author       = {{Synopsys}},
  title        = {Atomistic Simulation Software---{QuantumATK}},
  howpublished = {\url{https://www.synopsys.com/manufacturing/quantumatk.html}},
  note         = {Accessed 23 September 2026}
}

@MISC{sisl,
  author = {Papior, Nick},
  title = {{sisl}},
  year = {2023},
  doi = {10.5281/zenodo.597181},
  url = {https://doi.org/10.5281/zenodo.597181},
}

@ARTICLE{tbtrans,
  author = {Papior, Nick and Lorente, Nicol{\'a}s and Frederiksen, Thomas and
    Garc{\'i}a, Alberto and Brandbyge, Mads},
  title = {{I}mprovements on non-equilibrium and transport {G}reen function
    techniques: {T}he next-generation {T}ran{SIESTA}},
  journal = {Comput. Phys. Commun.},
  year = {2017},
  volume = {212},
  pages = {8--24},
  doi = {10.1016/j.cpc.2016.09.022},
}

\clearpage
\onecolumn
\thispagestyle{plain}

\begin{center}
    \LARGE \textbf{Supporting Information} \\[0.5em]

    \Large
    for \\[0.5em]

    \LARGE{\bf Conductance of silicon nanotube junctions \\in high magnetic fields}
\end{center}

\setcounter{section}{0}
\setcounter{subsection}{0}
\setcounter{figure}{0}
\setcounter{table}{0}

\renewcommand{\thesection}{S\arabic{section}}
\renewcommand{\thesubsection}{S\arabic{section}.\arabic{subsection}}

\renewcommand{\thefigure}{S\arabic{figure}}
\renewcommand{\thetable}{S\arabic{table}}

\section{DFT calculations}
The structural and electronic properties of the pristine (6,0) and (9,9) SiNTs were investigated using spin-polarised density functional theory (DFT) calculations within the generalised gradient approximation (GGA), employing the Perdew--Burke--Ernzerhof (PBE) exchange--correlation functional as implemented in the QuantumATK X-2025.06-SP1 package~\cite{Smidstrup2019,qATK}. Norm-conserving pseudopotentials from the PseudoDojo library were used together with the QuantumATK Medium basis set. The real-space mesh cut-off was set to 450\,Ry. The Brillouin zone was sampled using $1\times1\times63$ and $1\times1\times66$ $k$-point grids for the $(6,0)$ and $(9,9)$ SiNTs, respectively. The self-consistent-field calculations were converged using an energy tolerance of $10^{-5}$\,eV and a density-matrix tolerance of $10^{-4}$. Structural optimisations were continued until the maximum residual force acting on any atom was below $10^{-3}$\,eV/\AA\ and the maximum stress was below 0.1\,GPa. Following geometry optimisation, electronic band structures were calculated along the $\Gamma$--Z direction using 200 $k$-points.


\section{Tight binding parameters}


The TB parameters were obtained by fitting the band structure of SiNTs to the DFT band structure calculations described in Section S1. 
The fitting method consisted of optimizing the Slater-Koster parameters against the ab-initio bands ~\cite{Wang2021}. To describe the bands near the Fermi level and the metallicity of the nanotubes, the fit was restricted to the electronic states closest to the occupation boundary. For each nanotube, four bands below and four bands above the occupation boundary along the $\Gamma$–Z were used for the fit. For the $\tau$ nanotubes and $n$ fitted bands, the parameters were obtained by minimizing

\begin{equation}
\mathcal{L}(\mathbf{p})
=
\sum_{\tau}\frac{1}{N_{\tau}}
\sum_{n,k}
\left[
E_{\tau nk}^{\mathrm{TB}}(\mathbf{p})
+c_{\tau}
-E_{\tau nk}^{\mathrm{DFT}}
\right]^2
+
w^2\sum_{\tau}
\left[
g_{\tau}^{\mathrm{TB}}(\mathbf{p})
-g_{\tau}^{\mathrm{DFT}}
\right]^2
+
\frac{\lambda}{8}
\sum_{\alpha=1}^{8}
\left[
\frac{p_{\alpha}-p_{\alpha}^{(0)}}{p_{\alpha}}
\right]^2
\label{eq:tb-fitting-objective}
\end{equation}

where $c$ is the energy offset, $g$ is the valence band maximum / conduction band minimum, $p_\alpha$ the fitted Slater-Koster parameters. The residual weight is set to $w=2.5$ and the regularization parameter to $\lambda=10^{-3}$. All nanotubes were given the same weight and the minimisation was performed using bounded nonlinear least squares, with analytical eigenenergy derivatives obtained from the TB eigenvectors. At each optimisation step, the TB Hamiltonian 
was diagonalised and a rigid energy shift was calculated separately for each nanotube as the mean difference between the selected DFT and TB energies. To reduce the sensitivity to the initial parameter values, multiple randomly perturbed parameter sets were generated and optimized independently. The parameter set producing the lowest value of the objective function was selected as the final result.

\section{Additional results}

\begin{figure}[H]
     \centering
     \includegraphics[width=1.0\linewidth]{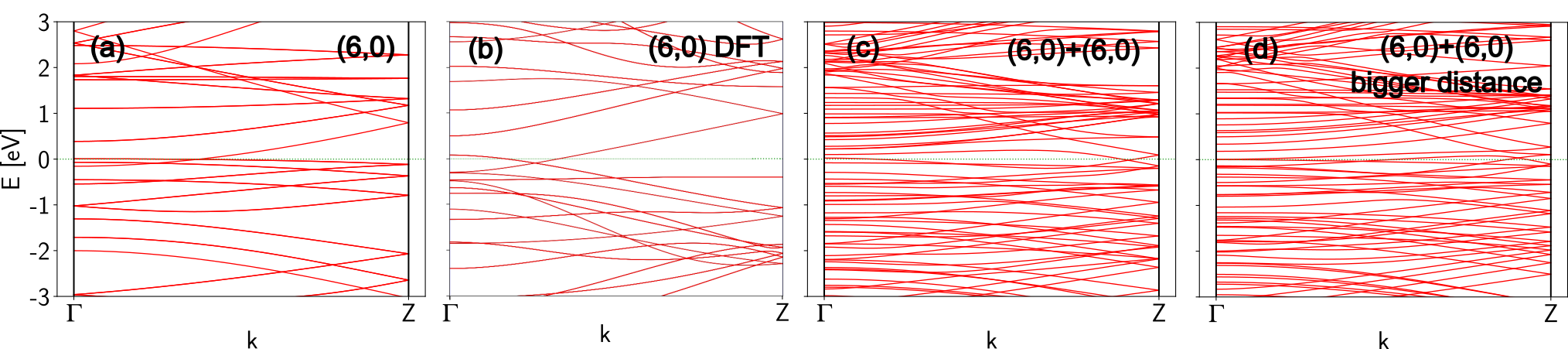}
     \caption{Band structures of (6,0) SiNT 2 unit cells obtained by (a) using our TB parameters and (b) DFT benchmarking. Band structures of (6,0)+(6,0) SiNT junction 2 unit cells obtained by using our TB parameters for inter-tube distances of (c) 0.40 nm and (d) 0.45 nm.}
     \label{fig:bands60}
\end{figure}

\begin{figure}[H]
     \centering
     \includegraphics[width=1.0\linewidth]{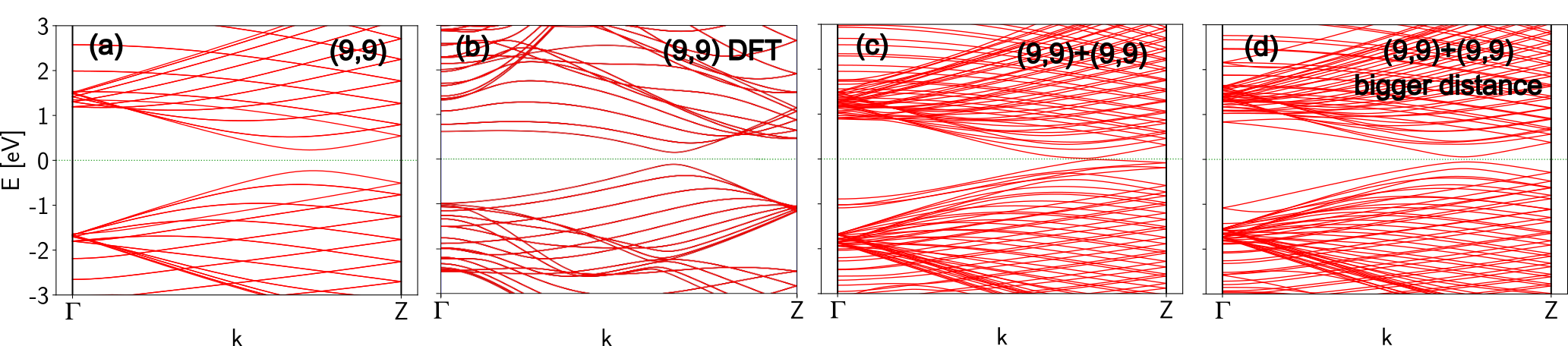}
     \caption{Band structures of (9,9) SiNT 2 unit cells obtained by (a) using our TB parameters and (b) DFT benchmarking. Band structures of (9,9)+(9,9) SiNT junction 2 unit cells obtained by using our TB parameters for inter-tube distances of (c) 0.40 nm and (d) 0.45 nm.}
     \label{fig:bands99}
\end{figure}

\begin{figure}[H]
     \centering
     \includegraphics[width=1.0\linewidth]{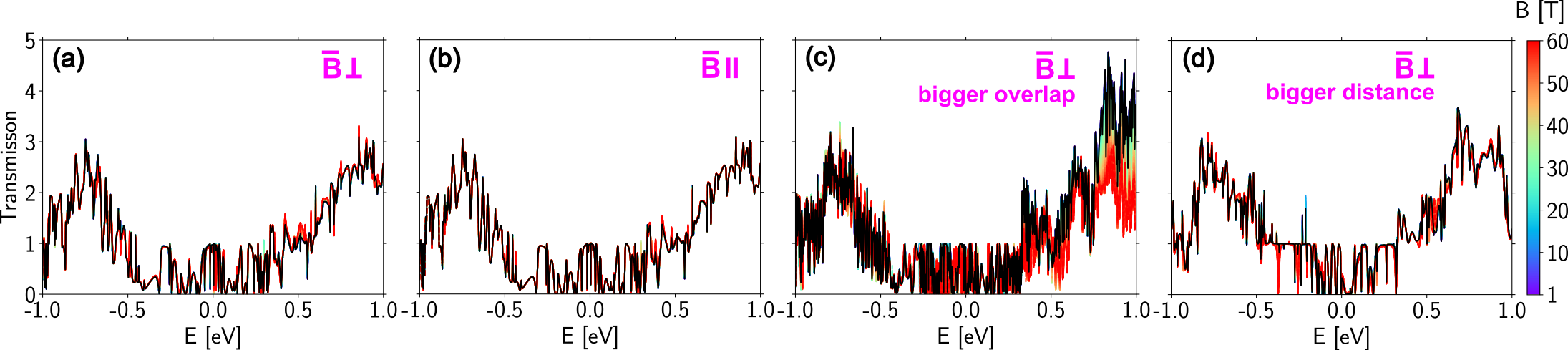}
     \caption{Transmission spectra in a bigger energy range of the $(6,0)+(6,0)$ SiNT junction at zero magnetic field (black lines) with an overlap lengths and inter-tube distances (a,b) 13.26 nm (20 units) and 0.40 nm, (c) 53.04 nm (80 units) and 0.40 nm, (d) 13.26 nm (20 units) and 0.45 nm, respectively. Coloured lines correspond to (a,c,d) perpendicular and (b) parallel magnetic fields up to 60 T.}
     \label{fig:fulltrans60}
\end{figure}

\begin{figure}[H]
     \centering
     \includegraphics[width=1.0\linewidth]{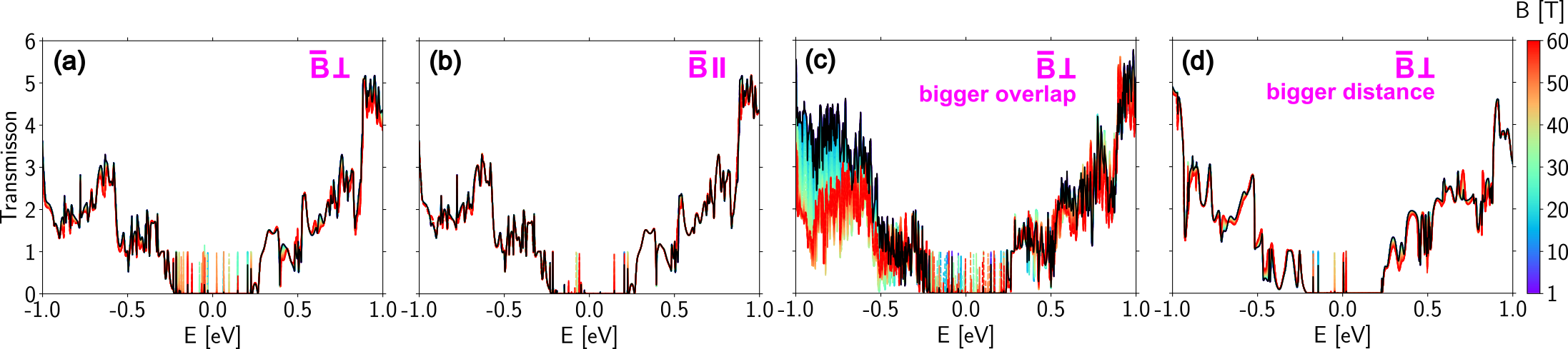}
     \caption{Transmission spectra in a bigger energy range of the $(9,9)+(9,9)$ SiNT junction at zero magnetic field (black lines) with an overlap lengths and inter-tube distances (a,b) 7.69 nm (20 units) and 0.40 nm, (c) 30.76 nm (80 units) and 0.40 nm, (d) 7.69 nm (20 units) and 0.45 nm, respectively. Coloured lines correspond to (a,c,d) perpendicular and (b) parallel magnetic fields up to 60 T.}
     \label{fig:fulltrans99}
\end{figure}

\end{document}